\documentclass[preprint,
 amsmath,amssymb,
 aps, physrev,
]{revtex4-2}
\usepackage{amssymb}
\usepackage[dvipsnames]{xcolor}
\usepackage{graphicx}
\usepackage[colorlinks=true, allcolors=blue]{hyperref}
\usepackage{dcolumn}
\usepackage{bm}

\begin{document}

\preprint{Number}

\title{Gradient catastrophe and Peregrine soliton in nonlinear flexible mechanical metamaterials}

\author{Antoine Demiquel}
 \email{Contact author: antoine.demiquel@insa-lyon.fr}
 \affiliation{Laboratoire d’Acoustique de l’Université du Mans (LAUM), UMR 6613, Institut d’Acoustique - Graduate School (IA-GS), CNRS, Le Mans Université, France}

\author{Vassos Achilleos}
\affiliation{Laboratoire d’Acoustique de l’Université du Mans (LAUM), UMR 6613, Institut d’Acoustique - Graduate School (IA-GS), CNRS, Le Mans Université, France}%

\author{Georgios Theocharis}
\affiliation{Laboratoire d’Acoustique de l’Université du Mans (LAUM), UMR 6613, Institut d’Acoustique - Graduate School (IA-GS), CNRS, Le Mans Université, France}%

\author{Vincent Tournat}%
 \email{Contact author: vincent.tournat@univ-lemans.fr}
\affiliation{Laboratoire d’Acoustique de l’Université du Mans (LAUM), UMR 6613, Institut d’Acoustique - Graduate School (IA-GS), CNRS, Le Mans Université, France}%

\date{\today}

\begin{abstract}
We explore the generation of extreme wave events in mechanical metamaterials using the regularization of the gradient catastrophe theory developed by A. Tovbis and M. Bertola for the nonlinear Schrödinger equation. According to this theory, Peregrine solitons can locally emerge in the semiclassical limit of the nonlinear Schrödinger equation. Our objective is to determine whether the phenomenon of gradient catastrophe can occur in a class of architected structures designated as flexible mechanical metamaterials, both with and without losses. We demonstrate theoretically and numerically that this phenomenon can occur in a canonical example of such flexible mechanical metamaterial, a chain of rotating units, studied earlier for its ability to support robust nonlinear waves such as elastic vector solitons. We find that in the presence of weak losses, the gradient catastrophe persists although the amplitude of extreme generated events is smaller and their onset is delayed compared to the lossless configuration.
\end{abstract}

\maketitle

\section{Introduction}
Rogue or freak waves are fascinating oceanic extreme phenomena that are rare and unpredictable. These giant waves emerge seemingly \textit{"from nowhere"} \cite{akhmediev_waves_2009}, towering over surrounding waves. Due to their rarity and the difficulty in predicting their occurrence, they are a threat to maritime safety and infrastructures \cite{muller_meeting_2015,bitner-gregersen_rogue_2015}. Before 1995, many scientists questioned the existence of these mysterious giant waves because of the lack of proof and explanation. When the first measurement of rogue waves was done on the Draupner platform in the North Sea, the scientific community increasingly started recognizing rogue waves as a real complex natural phenomenon. 

Studies have revealed that rogue waves are omnipresent \cite{dysthe_oceanic_2008,didenkulova_catalogue_2020} and that they can appear in different contexts induced by various linear and nonlinear mechanisms \cite{kharif_physical_2003,onorato_rogue_2013}. This diversity makes it challenging to establish a simplified definition of these unusual events \cite{akhmediev_editorial_2010}. For example, constructive interferences of wave groups \cite{gemmrich_generation_2022} is a linear mechanism that amplifies wave heights described by linear theories. Wave-current interactions as well as underwater topographical features, including shoals or deep canyons \cite{mallory_abnormal_1974,dysthe_refraction_2001,dysthe_oceanic_2008},
can also contribute to wave steepening leading to rogue wave formation. In addition, external factors such as atmospheric forcing \cite{janssen_interaction_2004,toffoli_observations_2024} can promote the growth of large wave groups. Wind applies pressure variations, transferring energy to the ocean surface. 
However, linear wave theory is limited in its ability to explain both the spontaneous occurrence and the extreme height of larger rogue waves. Indeed, these phenomena are much more frequent than the predicted wave height distribution \cite{longuethiggins_distribution_1980} (Rayleigh distribution). As a result, the study of nonlinear ocean dynamics has gained increasing popularity in understanding the stability and interactions of ocean waves which lead to the formation of coherent structures. It has been shown that ocean dynamics in the deep water limit are mathematically described by the nonlinear Schrödinger equation \cite{zakharov_stability_1968}.\\
This very same equation was found to describe accurately many other physical systems
\cite{ablowitz_discrete_2004,kivshar_optical_2003,peyrard_physics_2010}. This has led scientists to generalize the study of extreme wave events to many other physical disciplines, such as nonlinear optics \cite{kibler_peregrine_2010,copie_physics_2020}, Bose-Einstein condensates \cite{wen_matter_2011}, plasma physics \cite{bailung_observation_2011}, discrete systems as transmission lines \cite{remoissenet_waves_1999,ahmadou_dynamics_2023} or mechanical systems \cite{charalampidis_phononic_2018}.

The one-dimensional focusing NLS equation supports different dynamical mechanisms that lead to the emergence of coherent structures. For example, the modulation instability phenomenon \cite{demiquel_modulation_2023} originates from the exponential growth of the perturbations of an unstable plane wave background and can be triggered by either random or deterministic processes. Isolated coherent pulses and partially coherent fields \cite{el_koussaifi_spontaneous_2018} can exhibit self-focusing dynamics, as in the case of solitons on finite backgrounds or breathers. Amongst those, the Peregrine soliton (PS), localized both in time and space, may be appropriate to describe unique wave events \cite{peregrine_water_1983,akhmediev_rogue_2009,chabchoub_time-reversal_2014,dudley_rogue_2019}.

Recent mathematical proofs \cite{bertola_universality_2013} have shown that in the semiclassical limit of the 1D focusing NLSe, a universal mechanism regularizes the gradient catastrophe, leading to the local emergence of a Peregrine soliton. This phenomenon has already been experimentally observed in fiber optics \cite{tikan_universality_2017,tikan_local_2021} and water tanks \cite{tikan_prediction_2022}. 
Our goal is to leverage this knowledge in flexible mechanical metamaterials (FlexMM) to control the emergence of coherent structures in both time and space. Recent studies have shown that FlexMM are particularly well-suited for supporting and investigating nonlinear waves \cite{deng_nonlinear_2021}. This is due to the ability to harness their nonlinear dynamics through effective discrete mechanical models, while also taking advantage of recent advances in additive manufacturing, among other techniques, in order to conduct experimental tests \cite{deng_elastic_2017,deng_metamaterials_2018,deng_anomalous_2019,deng_propagation_2019}.

In Sec.~\ref{Sec:5.2}, we begin by presenting an overview of the considered FlexMM structure and its associated equations of motion, which include on-site dissipative terms. These terms describe linear dissipative mechanism where the amplitude at each discrete site decays over time at a rate proportional to its value. Using asymptotic expansion and multiple-scale methods as presented in \cite{demiquel_envelope_2024}. From the discrete equations of motion of the system, we derive an effective NLS equation (eNLS) that incorporates a linear loss term. This effective equation describes the slowly varying envelope of waves for the rotational degree of freedom (DOF) in the semi-discrete approximation. In Sec.~\ref{Sec:5.3}, the Peregrine soliton and the theoretical concept of gradient catastrophe regularization are presented in the semiclassical limit of the NLS equation. To continue, using an initial condition problem, we show that the dynamics of the NLS model are consistent with the dynamics of the FlexMM with and without losses, in Sec.~\ref{Sec:5.5} and Sec.~\ref{Sec:5.6} respectively. Finally, Sec.~\ref{sec:smaller_lattice_system} considers a smaller lattice system model, paving the way for the experimental study of the above described phenomena.

\section{Modulated waves in FlexMM in the weakly nonlinear and dissipative regime}
\label{Sec:5.2}

\begin{figure}
    \centering
    \includegraphics[width=1\linewidth]{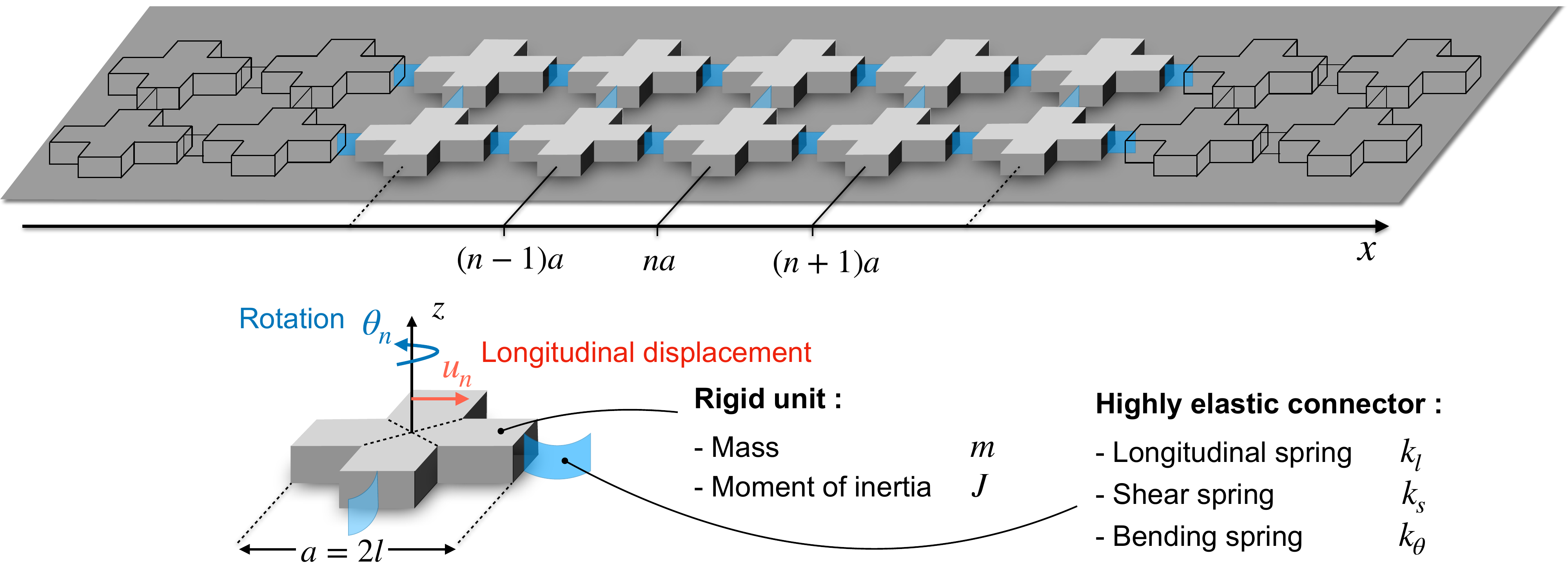}
    \caption{Outline of the considered FlexMM composed of two rows of rigid mass units (gray crosses) connected by highly elastic connectors (blue areas) extending along the x-direction with an infinite periodic configuration. The rigid units are characterized by a mass $m$ and a moment of inertia $J$. The elastic connectors are characterized by effective stiffnesses $k_l$, $k_s$ and $k_\theta$. Each pair of unit $n$ can rotate ($\theta_n$) and longitudinally translate ($u_n$) from its equilibrium position.}
    \label{System_schematic}
\end{figure}

In the low-frequency regime, the considered FlexMM structure can be approximated using a lumped-element model \cite{demiquel_envelope_2024}, represented as a one-dimensional chain composed of two lines of rigid units with mass $m$ and moment of inertia $J$. These units are periodically connected by highly elastic connectors characterized by three effective massless springs: a longitudinal spring with stiffness $k_l$, a shear spring with stiffness $k_s$, and a bending spring with stiffness $k_\theta$. For simplicity, only symmetric motion relative to the axis of symmetry between the two lines is considered. Under this assumption, each mass in the system has two degrees of freedom: a longitudinal displacement $u$ and a rotational motion $\theta$, where the rotation angle is defined as positive in the counterclockwise (trigonometric) direction. The system is outlined in Fig.~\ref{System_schematic}.

Under the above assumptions, the normalized equations of motion of the system take the following form,

\begin{subequations}
 \begin{align}
 \frac{d^2U_n}{dT^2} &= U_{n+1}-2U_n+U_{n-1}-\frac{\cos\theta_{n+1}-\cos\theta_{n-1}}{2}-\Gamma_U\frac{dU_n}{dT} \,,\\
\frac{ 1}{\alpha^2}\frac{d^2\theta_n}{dT^2} &= K_\theta\left(\theta_{n-1}-4\theta_n+\theta_{n+1}\right)- K_s \cos\theta_n\left[\sin\theta_{n-1} + 2\sin\theta_n+\sin\theta_{n+1}\right]
\nonumber\\&
- \sin\theta_n \left[2(U_{n+1}-U_{n-1})+4-\cos\theta_{n-1}-2\cos\theta_n-\cos\theta_{n+1}\right]-\Gamma_\theta\frac{d \theta_n}{dT}\,.
\label{Eq_theta}
 \end{align}
 \label{Full_discrete_chap_5}
\end{subequations}

The following normalized variables and parameters are defined: the longitudinal displacement of unit $n$ as $U_n = u_n / a$, the normalized time as $T = t\sqrt{k_l / m}$, an inertial parameter $\alpha = a\sqrt{m / (4J)}$, and the stiffness parameters $K_\theta = 4k_\theta / (k_l a^2)$ and $K_s = k_s / k_l$. The parameters chosen in this work are those used in our previous research \cite{demiquel_envelope_2024}.
They are the following,
\begin{equation}
 \alpha=1.815, \hspace{0.5cm }K_s=0.1851, \hspace{0.5cm} K_\theta=1.534\text{e}^{-2},
 \label{parameters}
 \end{equation}
where the value of $\alpha$ originates from the experimental setup constituted of Lego\textsuperscript{\textregistered} bricks and used in \cite{deng_metamaterials_2018}. 

In Eq.~(\ref{Full_discrete_chap_5}), dissipation is accounted for via linear damping associated with the respective translation ($\Gamma_U$) and rotation ($\Gamma_\theta $) motions of each unit $n$. Viscous-type damping is a commonly used model in discrete lattice systems to describe local energy dissipation \cite{peyrard_physics_2010}. For example it has been recently used to model loss effects on propagating unidirectional dissipative solitons in one-dimensional active flexible mechanical metamaterials \cite{brandenbourger_non-reciprocal_2024}, where each unit can only rotate.

\subsection{Effective nonlinear Schrödinger equation with linear loss term}
In the weakly nonlinear and dissipative regime we develop a theoretical model in order to describe the propagation of slow-modulated traveling waves. We assume weak dissipation where the damping parameters are of order $\epsilon^2$,
\begin{equation}
        \Gamma_{U/\theta} =\epsilon^2 \gamma_{U/\theta}\,,
\end{equation}
with $\epsilon \ll 1$ a small parameter. In this regime, the discrete dispersion relations in the linear limit show almost no difference between the lossy and lossless cases (with $\Gamma_U = \Gamma_\theta = 0$), as the losses are sufficiently small. The corresponding dispersion relations are expressed as follows,
 \begin{subequations}
\begin{align}
 \omega^{(1)} & =2\sin\left(\frac{k}{2}\right)\, ,\label{omega_1:subeq1_chap_5}\\
 \omega^{(2)}&=\pm\sqrt{4\alpha^2(K_s-K_\theta)\cos^2\left(\frac{k}{2}\right)+6\alpha^2 K_\theta} \, .\label{omega_2:subeq2_chap_5}
 \end{align}
 \label{dispersion_relation}
\end{subequations}

We focus on weakly nonlinear solutions and consequently substitute the following expansions ,
\begin{equation}
 \cos{\theta_n} = 1-\frac{\theta_n^2}{2}+\ldots \, , \hspace{0.5cm}
 \sin{\theta_n} =\theta_n-\frac{\theta_n^3}{6}+\ldots \, ,
\end{equation}
to Eq.~(\ref{Full_discrete_chap_5}). By keeping terms up to the cubic order we end up with the following set of equations of motion,
\begin{subequations}
 \begin{align}
 &\frac{d^{2} U_{n}}{ dT^{2}} =\; U_{n+1}-2 U_{n}+U_{n-1}-\frac{\theta_{n-1}^2 -\theta_{n+1}^2}{4}-\epsilon^2\gamma_U\frac{d U_n}{dT}\label{sub:Motion_equation_Order_3_NL_U}\, ,\\
 &\frac{d^{2} \theta_{n}}{dT^{2}} =-\alpha^2\left(K_s-K_{\theta}\right)\left(\theta_{n-1}+2 \theta_{n}+\theta_{n+1}\right)-6K_\theta\alpha^2\theta_n+\alpha^2(K_s-1)\theta_n^3\nonumber\\
 &-\alpha^2\frac{\theta_n}{2}\left(\theta_{n-1}^2+\theta_{n+1}^2\right)+\frac{\alpha^2 K_s}{6}\left(\theta_{n-1}^3+2\theta_n^3+\theta_{n+1}^3\right)+K_s\alpha^2 \frac{\theta_n^2}{2}\left(\theta_{n-1}+\theta_{n+1}\right)\nonumber\\
 &-2\alpha^2\theta_n(U_{n+1}-U_{n-1})-\epsilon^2\alpha^2\gamma_\theta\frac{d \theta_n}{dT}\label{sub:Motion_equation_Order_3_NL_Theta} \,.
 \end{align}
 \label{Motion_equation_Order_3_NL}
\end{subequations}

To study modulated traveling waves, we use the semi-discrete approximation \cite{kivshar_modulational_1992,daumont_modulational_1997,remoissenet_low-amplitude_1986}, where a
carrier wave, governed by the discrete dispersion relation, is
modulated by a slowly varying envelope function $F_{1,n}$ treated
in the continuum limit as $F_1$. Given the presence of quadratic terms $\sim \theta^2$ in Eq.~(\ref{Motion_equation_Order_3_NL}), we include both a dc-term $G_{0,n}$ and a term $G_{2,n}$ that oscillates with a $2\sigma_n$ phase. Hence, we seek solutions of the following form,
\begin{subequations}
\begin{align}
U_n &= \epsilon G_{0,n}(T)+\epsilon^2\left[G_{2,n}(T)e^{2i\sigma_n}+G_{2,n}^*(T)e^{-2i\sigma_n}\right] \, ,\\
\theta_n &= \epsilon\left[F_{1,n}(T)e^{i\sigma_n}+F_{1,n}^*(T)e^{-i\sigma_n}\right] \, ,
\end{align}
\label{perturbative_expansion_chap_5}
\end{subequations}
where the fast oscillations of the carrier wave of phase $\sigma_n =kn-\omega T$ are chosen to obey the discrete dispersion relation in Eq.~(\ref{omega_2:subeq2_chap_5}). Using a multiple scales method \cite{peyrard_physics_2010,holmes_introduction_1995}, we obtain a hierarchy of equations at various orders in $\epsilon$ as detailed in appendix \ref{appendix_A}. The envelope of the modulated wave $\theta_n$ follows a nonlinear Schrödinger equation incorporating a linear loss term,
\begin{equation}
 i\frac{\partial F_1}{\partial \tau_2}+ P\frac{\partial^2 F_1}{\partial \xi_1^2}+Q|F_1|^2F_1=-i\frac{\gamma_\theta \alpha^2}{2}F_1\, ,
 \label{NLS_PQ_chap5}
 \end{equation}
 with, 
\begin{subequations}
 \begin{align}
 P =& \frac{\alpha^2(K_\theta-K_s)\cos(k)-v_g^2}{2\omega}\, ,\\
 Q =& \frac{1}{2\omega}\left[8K_s\alpha^2\cos^2\left(\frac{k}{2}\right)-\alpha^2(5+\cos(2k))+\frac{\alpha^2\sin^2(2k)}{2\left(\sin^2(k)-\omega^2\right)}-\frac{4\alpha^2}{v_g^2-1}\right] \, ,
 \end{align}
 \label{PQ_chap_5}
\end{subequations}

where $P$ and $Q$ are coefficients that depend on the structure geometry and the wave number. For more details, the analysis of the focusing (respectively defocusing) region has been displayed in Fig.~3(a) of \cite{demiquel_envelope_2024}. For this study, we fix the wave number to $k=0$. Using the virtual FlexMM parameters in Eq.~(\ref{parameters}) and $k=0$, the dispersive coefficient (nonlinear coefficient) $P$ ($Q$) is negative. As a consequence the resulting effective NLS equation is focusing ($PQ>0$). $G_0$ and $G_2$ can be obtained from $F_1$,
\begin{subequations}
\begin{align}
 G_0(\xi_1,\tau_2) &= \frac{1}{v_g^2-1}\int|F_1(\xi_1,\tau_2)|^2 d\xi_1 \,,\label{G0_chap_5}\\ 
 G_2(\xi_1,\tau_2) &= \frac{i\sin(2k)}{8\left(\sin^2(k)-\omega^2\right)}F_1(\xi_1,\tau_2)^2\,. 
 \end{align}
 \label{G_0,G_2}
\end{subequations}
$\xi_1 =\epsilon(X-v_g T)$ and $\tau_2=\epsilon^2 T$ are the time and space scales used for the expansion. It is interesting to note that aside from the linear loss term in Eq.(\ref{NLS_PQ_chap5}), the coefficients $P$ and $Q$ in Eq.(\ref{PQ_chap_5}) as well as the solutions $G_0$ and $G_2$ in Eq.~(\ref{G_0,G_2}) are identical to those of the lossless model previously developed \cite{demiquel_envelope_2024}. In conclusion, in the weakly nonlinear regime, only $\Gamma_\theta$ impacts the dynamics of the system through its role on the envelope of the modulated wave $\theta_n$ at the leading order.

\section{A review of the gradient catastrophe phenomenon of the NLSE}
\label{Sec:5.3}
We know that the evolution of solitons can be robust thanks to a perfect balance between nonlinearity (which tends, for example, to concentrate a wave impulse by forming a shock) and dispersion (which tends to spread the same impulse).  When the nonlinearity effect is much stronger than the dispersion one, the wave behavior is dominated by nonlinear effects resulting in wave compression and breather solutions emergence  \cite{Mollenauer:83}. In this section, we aim to determine the conditions both in terms of structural configuration and excitation parameters under which such solutions can emerge in FlexMMs.

\subsection{Semiclassical limit}
We now review the semiclassical limit of the NLS equation to subsequently apply it to the effective NLS model of FlexMM obtained in Eq.~(\ref{NLS_PQ_chap5}). The semiclassical limit of the NLS equation, also known as the zero dispersion limit, corresponds to the strongly nonlinear regime obtained when the dispersion effects are much weaker than the nonlinearity effects.
For the semiclassical limit analysis, the NLS equation, cf.~Eq.~(\ref{NLS_PQ_chap5}), can be written in the following form,

\begin{equation}
 i\frac{\partial \psi}{\partial \mathcal{T}}+\frac{1}{2N} \frac{\partial^2 \psi}{\partial \mathcal{X}^2}+N|\psi|^2\psi=-i\mathcal{G}_\theta \psi \, ,
 \label{NLS_semi_classique}
\end{equation}
where $N$ is the soliton number (see below) and
$\psi$, $\mathcal{X}$ and $\mathcal{T}$ are given by the following normalization relations,

\begin{equation}
 \psi =\frac{F_1}{A_0} \,,\hspace{0.5cm}
\mathcal{T} =\frac{\tau_2}{\sqrt{\tau_{NL}\tau_D}}\,,\hspace{0.5cm}
 \mathcal{X} =\frac{\xi_1}{L_e}\,.
 \label{change_of_variables_chap_5}
\end{equation}
The dissipation coefficient of Eq.~(\ref{NLS_semi_classique}) is connected to that of the FlexMM equations of motion through the following relationship,
\begin{equation}
    \mathcal{G}_\theta= \frac{\gamma_\theta}{2}\frac{\alpha^2 N}{|Q| A_0^2}\,.
    \label{mathcal_G_theta}
    \end{equation}
    
The soliton number $N>0$ controls the dispersion to nonlinearity ratio,
\begin{equation}
 N=\sqrt{\frac{\tau_D}{\tau_{NL}}}\,,
 \label{number_of_soliton}
\end{equation}
where $\tau_D = \frac{L_e^2}{2|P|}$ and $\tau_{NL} = \frac{1}{|Q|A_0^2}$ are the characteristic dispersive and nonlinear times (at the order of the $\tau_2$ time scale) defined by the amplitude $A_0$ and width $L_e$ of the initial pulse. It is worth noting that $N$ may take non-integer values.\\
When $N$ is an integer, an $N$-soliton solution \cite{yang_nonlinear_2010} of Eq.~(\ref{NLS_semi_classique}) at $\mathcal{T}=0$ gives,
\begin{equation}
 \psi(\mathcal{X},0) = \text{sech}\left(\mathcal{X}\right)\,.
\end{equation}
These solitons exhibit more complex dynamics than the fundamental 1-soliton, including periodic oscillations called breathers or multi-peak structures produced by the interaction and the superposition of multiple fundamental solitons, explained by the Inverse Scattering Transform method \cite{yang_nonlinear_2010}. In the semiclassical limit of the one-dimensional focusing NLS equation, 
recent studies \cite{tikan_universality_2017,tikan_local_2021,tikan_nonlinear_2022} revealed that another fundamental mechanism, called gradient catastrophe, leads to the emergence of localized structures using the self-focusing property of the NLS equation. These localized structures have been proven to asymptotically  approach the Peregrine soliton when $N\rightarrow \infty$ \cite{bertola_universality_2013,tikan_universality_2017}.

\subsection{Peregrine soliton of the NLSE}
\begin{figure*}[ht!]
 \centering
\includegraphics[width=0.9\linewidth]{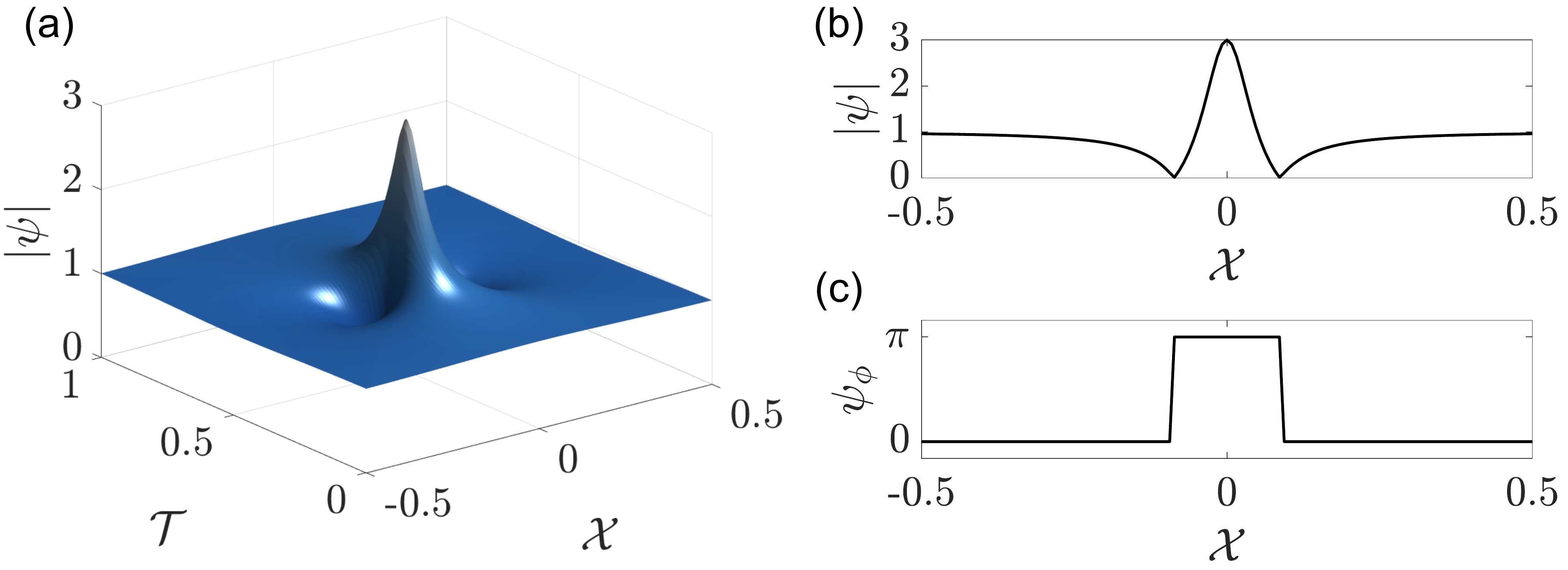}
 \caption{Analytical Peregrine soliton solution of Eq.~(\ref{NLS_semi_classique}), for $N=10$ and $a_0=1$. In panel (a), a spatiotemporal representation is displayed while in panels (b-c), a spatial profile at the maximum compression moment $\mathcal{T}=\mathcal{T}_m=0.5$ of the amplitude and phase is visible.}
 \label{Peregrine_soliton_th}
\end{figure*}

The Peregrine soliton (PS) is a rational solution of the NLS equation originally proposed in 1983 by D.H. Peregrine \cite{peregrine_water_1983}. It is a solution of high amplitude with a stiff wavefront profile localized both in time and space. Mathematically, it is a solution of Eq.~(\ref{NLS_semi_classique}) with no dissipation ($\mathcal{G}_\theta =0$) that writes,
\begin{equation}
\psi(\mathcal{X},\mathcal{T}) =a_0\frac{1-4[1+2ia_0^2N(\mathcal{T}-\mathcal{T}_m)]}{1+4a_0^2N^2\mathcal{X}^2+4a_0^4N^2(\mathcal{T}-\mathcal{T}_m)^2}e^{ia_0^2N(\mathcal{T}-\mathcal{T}_m)} \, ,
 \label{Peregrine_sol}
\end{equation}

where $a_0$ is the continuous background amplitude and $\mathcal{T}_m$ is the moment of maximum compression. The analytical PS of \,Eq.~(\ref{Peregrine_sol}) is displayed in Fig.~\ref{Peregrine_soliton_th}.

At the maximum compression moment visible in Fig.~\ref{Peregrine_soliton_th} (b), the amplitude of the PS reaches three times the continuous background amplitude $\psi(0,\mathcal{T}_m)= 3a_0$ including a $\pi$-~phase jump, cf. Fig.~\ref{Peregrine_soliton_th} (c). After this point, the amplitude of the solution decreases, expands, and finally disappears, cf. Fig.~\ref{Peregrine_soliton_th} (a). Its shape and localization in both space and time resemble the sudden and transient nature of rogue waves \cite{shrira_what_2010}. This is why the Peregrine soliton is often considered as a possible explanation of this kind of extreme wave event.

\subsection{Local emergence of the Peregrine soliton by the regularization of the gradient catastrophe}
\label{local_e_P_s}
In the semiclassical limit of the one-dimensional focusing nonlinear Schrödinger equation without linear loss term ($\mathcal{G}_\theta = 0$) and for $N\gg 1$, it has been established \cite{tikan_universality_2017,copie_physics_2020} that during the initial stages of its evolution, a wide pulse undergoing propagation is primarily influenced by nonlinear effects while dispersive effects are negligible. This dominance of nonlinearity leads to a self-steepening of both the phase and amplitude profiles of the pulse, reaching a critical point ($\mathcal{T}_c$, $\mathcal{X}_c$) where the derivatives of phase and amplitude become infinite. At this point, the gradient catastrophe phenomenon emerges and the pulse becomes localized at position $\mathcal{X}_c$ and moment $\mathcal{T}_c$. Following the gradient catastrophe occurrence ($\mathcal{T}>\mathcal{T}_c$), dispersive effects cannot be ignored, leading to the regularization of the gradient catastrophe through the emergence of localized breathers. It has been shown in \cite{bertola_universality_2013} that the maximum compression point occurs at time,
\begin{equation}
 \mathcal{T}_m = \mathcal{T}_c+C N^{-4/5}\, ,
 \label{T_m}
\end{equation}
with $C$ a universal constant defined as,
\begin{equation}
 C =2.38\left(\frac{5|C_1|}{4}\right)^{1/5}(2b_0)^{-3/2}\left(1+\mathcal{O}\left(N^{-4/5}\right)\right) \, .
\end{equation}
When the initial pulse is a $N$-soliton solution,
\begin{equation}
 \mathcal{T}_c = 1/2 \, ,\hspace{0.5cm} \, b_0 = \sqrt{2} \, ,\hspace{0.5cm} C_1 = \frac{32\sqrt{2i}}{15\times 2^{9/4}} \, .
\end{equation}
The first localized structure that emerges can be asymptotically approximated by a Peregrine soliton of the NLSE, cf. Eq.~(\ref{Peregrine_sol}). 
For $N\rightarrow \infty$, the amplitude of the Peregrine soliton reaches the asymptotic limit $\psi(\xi,\mathcal{T}_m) = 3\sqrt{2}$. 

\section{Gradient catastrophe in the FlexMM: lossless configuration}
\label{Sec:5.5}
The theoretical concept and key results for the gradient catastrophe regularization have been presented above. We would now like to find out whether such a phenomenon can be observed in FlexMM structures. From the effective NLS equation of the architected structure we can identify the appropriate initial conditions for the FlexMM that lead to gradient catastrophe. In the next sections, we compare the dynamics of the FlexMM with the effective NLS equation evolution. To this end, we carry out a series of simulations on both the NLS equation (\ref{NLS_semi_classique}) and the FlexMM discrete-lattice equations of motion (\ref{Full_discrete_chap_5}). As the phenomenon is known for the NLS equation, the aim here is to vary the number of $N$-soliton solutions to numerically prove that the first emerged localized structure in the rotational degree of freedom evolution of the FlexMM can also be well fitted by a Peregrine soliton.

\subsection{Numerical integration of the NLS equation and FlexMM equations of motion 
}
\label{sec:Numerical_NLS_chap_5}
On the one hand, to numerically solve the NLS equation Eq.~(\ref{NLS_semi_classique}), we use the exponential time difference fourth-order Runge-Kutta (ETDRK4) scheme \cite{cox_exponential_2002,kassam_fourth-order_2005}. It is a powerful integration scheme that provides a robust, accurate, and efficient solution for stiff differential equations by exactly handling the stiffness arising from the linear components of the system. The integration is done using an $N$-soliton exact solution as initial condition,
\begin{equation}
 \psi(\mathcal{X},\mathcal{T}=0) = \text{sech}\left(\mathcal{X}\right)~.
 \label{IC_NLS_chap_5}
\end{equation}

Direct numerical simulations of the discrete set of equations (\ref{Full_discrete_chap_5}) are employed to validate the analytical predictions and to compare them to the NLS equation evolution. 

On the other hand, the FlexMM system (\ref{Full_discrete_chap_5}) is solved using a fourth-order Runge-Kutta iterative integration scheme with free boundary conditions at both ends. We use the $N$-soliton solution cf. Eq.~(\ref{IC_NLS_chap_5}) as initial condition. The latter can be expressed as a function of
$(\tau_2,\xi_1)$,
 \begin{equation}
 F_1(\xi_1,0) = A_0\text{sech}\left(\frac{\xi_1}{L_e}\right)~, 
 \label{F1_chap_5}
\end{equation}
with 
\begin{equation}
 L_e = \frac{N}{A_0}\sqrt{\frac{2|P|}{|Q|}}~,
 \label{Le}
\end{equation}
where $A_0$ is the amplitude of the pulse in the co-moving frame coordinate system and where $\xi_1=\epsilon(X-v_gT)$.

To determine the initial conditions for the FlexMM dynamical equations, namely the corresponding lattice waves for $\theta$ and $U$ fields, using Eq.~(\ref{perturbative_expansion_chap_5}), we obtain for the rotation,
\begin{equation}
 \theta_1(X,0)= 2 A_0 \text{sech} \left[\frac{\epsilon}{L_e}(X-X_0)\right]\cos(kX) \,.
 \label{Analy_theta_chap_5}
\end{equation}
 $k$ is the wave number of the carrier wave following the dispersion relation of the $\theta$ branch, see ~Eq.~(\ref{omega_2:subeq2_chap_5}). The combination of Eqs.~(\ref{F1_chap_5}) and (\ref{G0_chap_5}) gives the following expression for the dc-term initial condition,
\begin{equation}
 U_0(X,0) =\frac{A_0^2 L_e}{v_g^2-1} \tanh\left[\frac{\epsilon (X-X_0)}{L_e}\right]\,.
 \label{Analy_U_chap_5}
\end{equation}
The initial conditions on the velocity $\dot{\theta}_1(X,0)$ and $\dot{U}_0(X,0)$ are deduced from the analytical expression of the BEVS \cite{demiquel_envelope_2024} [see Eqs.~(\ref{Analy_theta_appendix_B}-\ref{Analy_U_appendix_B}) in appendix~\ref{appendix_B}]. Since the solution only is taken into account up to the first order of perturbation ($\epsilon)$, the initial conditions write, 
\begin{equation}
\begin{split} &\theta(X,0)= \epsilon \theta_1(X,0)\, ,\hspace{0.5cm}
\dot \theta(X,0) = \epsilon\dot \theta_1(X,0)\, , 
\end{split}
 \label{IC_T_chap_5}
 \end{equation}
 \begin{equation}
\begin{split} &U(X,0) = \epsilon U_0(X,0) \, ,\hspace{0.5cm}
\dot U(X,0) = \epsilon \dot U_0(X,0)\, .
 \end{split} 
 \label{IC_U_chap_5}
 \end{equation}
For the simulations, the initial conditions are centered at $X_0$, the middle of the lattice. The parameters used to run the simulations are the following. The integration of the equations of motion is done on $n_s=2000$ sites over a total time of $T_f = 10^4$, with a time step $dt=0.05$. The initial amplitude is set to $A_0=5$, the wave number of the carrier wave is $k=0$, and $\epsilon=0.01$. For the NLS equation, the ETDRK4 scheme integrates over a time $\mathcal{T}_f=\epsilon^2 T_f\sqrt{T_{NL} T_D} $,
with a time step of $h=1\text{e}^{-3}$. The spacial domain is a grid of size $L_x=\epsilon n_s/L_e$, divided in $N_x =2^{10} $ points.

We choose to use a large number of sites ($n_s=2000$) to solve the equations of motion to ensure the validity of the NLS model and to allow analysis of the system's dynamics with a large number of solitons ($N$). The analysis of the system with a reduced number of sites is explored in Sec.~\ref{sec:smaller_lattice_system}.

\subsection{Higher-order solitons in FlexMM}
\label{Numerical_results_chap_5}
\begin{figure*}[ht!]
 \centering
\includegraphics[width=1\linewidth]{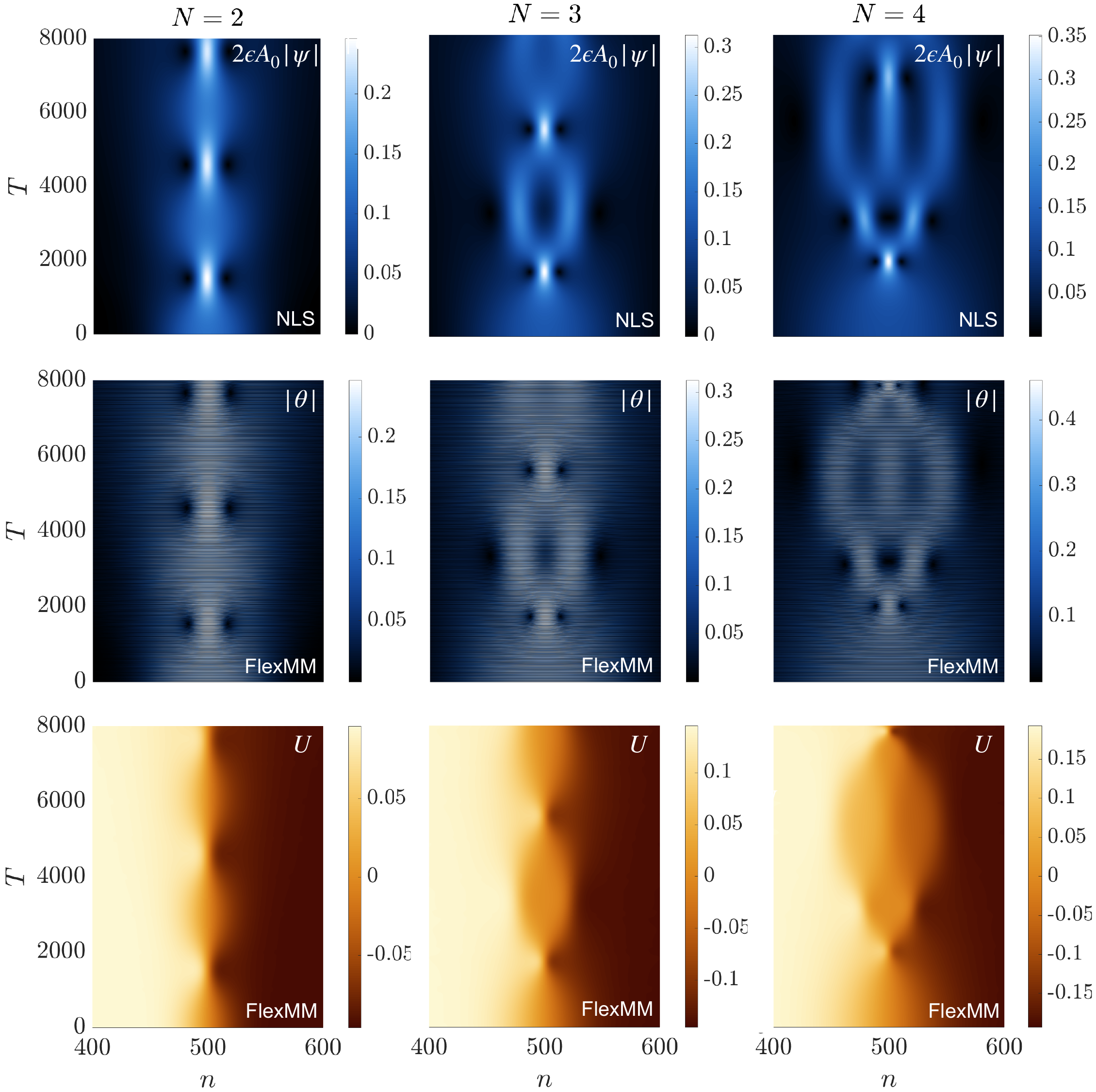}
\vspace{-0.7cm}
\caption{Top line: numerical solutions of the NLS equation. Middle and bottom lines: numerical solutions of the
rotational  ($\theta$) and  the longitudinal ($U$) degrees of freedom of the 
FlexMM, respectively. Each column corresponds to a specific number of solitons $N=\{2,3,4\}$.} 
 \label{Higher_orders}
\end{figure*}

We now focus on the higher-order soliton solutions of FlexMMs predicted by the effective NLS equation. In the first row of Fig.~\ref{Higher_orders}, we display the results of the numerical integration of the NLS equation Eq.~(\ref{NLS_semi_classique}) while in the second (third) row, we show the numerical results of the lattice equations Eq.~(\ref{Full_discrete_chap_5}) for the rotational (longitudinal) component, choosing $N=\{2, 3, 4\}$.

Let us first discuss the $N=2$ soliton solution. It corresponds to the first column of Fig.~\ref{Higher_orders}. Concerning the NLSE, this solution consists of the nonlinear superposition of two fundamental, non-traveling solitons. Together, they form a bound state that results in a breathing solution characterized by a temporally-periodic evolution. 

Concerning the FlexMM discrete lattice model, we observe that the envelope of the $\theta$ component follows the NLS equation evolution with great accuracy. In addition, due to the nonlinear coupling with the $U$ component, one can observe a corresponding pattern in the evolution of the longitudinal motion is driven by the rotational motion.

Similarly, for higher order solitons, for example $N=3$ and ($N=4$), the emergence of localized breathing structures is induced by the interaction of three (four) bright solitons as shown in the second (third) column of Fig.~\ref{Higher_orders}.
The good agreement between the FlexMM discrete evolution and the NLS evolution suggests that new non-traveling 2-components envelope vector solitons can emerge in the lattice and are well predicted by the effective nonlinear Schrödinger equation. The rotational field $\theta$ produces high-order breather solitons and couples to breathing kink-like profiles arising from the longitudinal displacement $U$. These vectorial solutions imply that a local rotation of the chain elements results in their contraction along the $x$-axis.  

It is essential to emphasize that the vectorial nature of these solutions plays a key role in their stability during propagation. Indeed, the $Q$ coefficient of the effective NLS equation is modified due to the coupling between the DOFs ($U$,$\theta$), as shown in Appendix A. Moreover, the study conducted on the same system for the modulation instability phenomenon \cite{demiquel_modulation_2023} shows that the presence or absence of this nonlinear coupling can lead to two distinct dynamic behaviors. This contrast highlights the crucial impact of interactions between the DOFs on the evolution of vectorial solutions or their robustness.

\subsection{Gradient catastrophe in FlexMM}

\begin{figure*}[ht!]
 \centering
\includegraphics[width=1\linewidth]{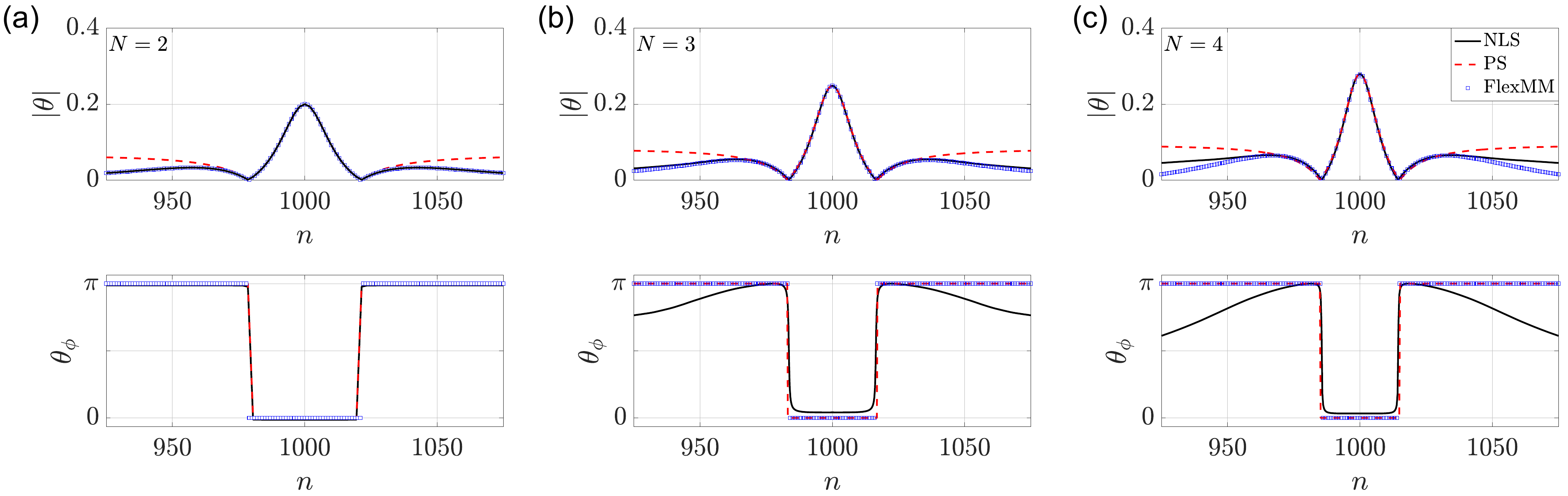}
 \caption{Profile and phase of the first localized structure for $N=2$ in panel (a), $N=3$ in panel (b), and $N=4$ in panel (c) at the maximum compression moment for the FlexMM (blue square) and the NLS equation (black line). For each $N$, the coherent structure is compared to a scaled PS (red dashed line).}
\label{Higher_orders_soliton_maximum_compression}
\end{figure*}

As noted in Sec.~\ref{local_e_P_s}, the gradient catastrophe phenomenon is regularized by the local emergence (at $T=T_m$) of a Peregrine soliton as an asymptotic solution of the focusing NLS equation. This property is verified in Fig.~\ref{Higher_orders_soliton_maximum_compression}. 

In this figure, the discrete spatial profile at the maximum compression moment for the rotational motion of the FlexMM is plotted (blue squares) and compared to the NLSE prediction (black line). The analytic PS (dashed red line) agrees with the discrete simulations for both the spatial and phase profiles.

We note that we determined the PS continuous background $a_0$, cf.~Eq.~(\ref{Peregrine_soliton_th}), from the maximum amplitude of the rotational component (NLSE simulation). In this case $a_0=~|\theta(X_0,T_m)|/3$.

\begin{figure}[ht!]
 \centering
\includegraphics[width=1\linewidth]{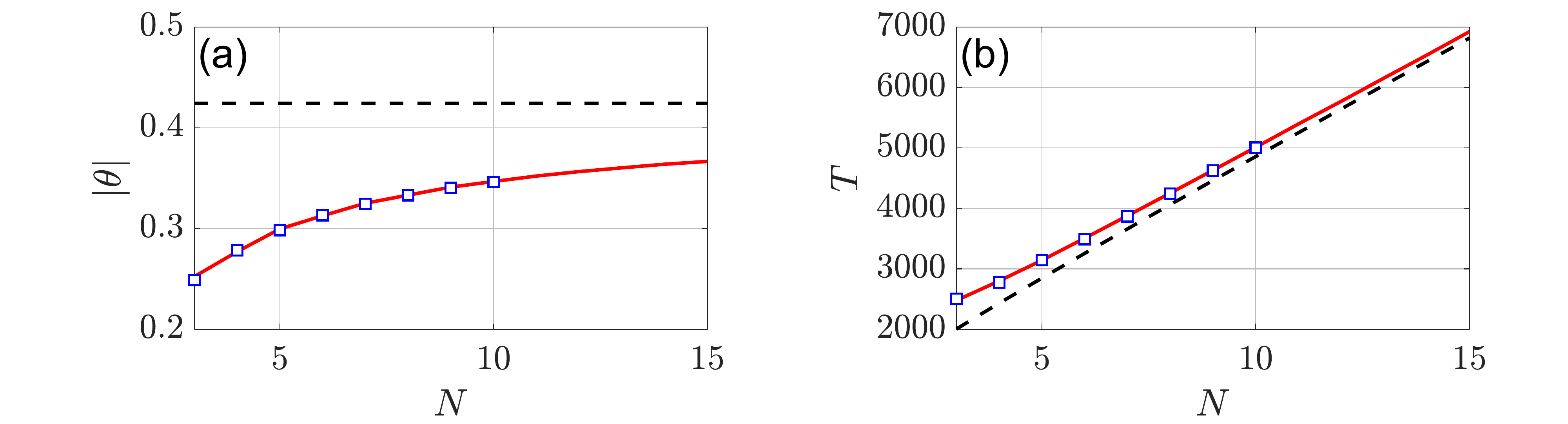}
 \caption{Numerical simulations displaying the maximum amplitude (a) of the rotational evolution at the compression moment (b) as a function of $N$. The black dashed line represents the theoretical predictions from Eqs.~(\ref{asymptotic}-\ref{T_m}). The red line represents the simulation results using the NLS equation, and the blue squares represent those of the FlexMM.}
\label{Max_Amp_position}
\end{figure}

For a more detailed analysis of the theoretical prediction accuracy, we show the amplitude of the first localized structure in Fig.~\ref{Max_Amp_position}(a) and the maximum compression moment in Fig.~\ref{Max_Amp_position}(b) for a wide range of $N$ values. For the discrete equations of the FlexMM, $N$ takes values from 3 to 10 for which it is clear that the FlexMM and the NLS model are in excellent agreement. If for $N=10$, the numerically calculated value of $T_m$ is close to the theoretical estimation, we observe that the amplitude remains significantly far from the asymptotic value,
\begin{equation}
 \theta_{N\rightarrow \infty} = 6\sqrt{2}\epsilon A_0.
 \label{asymptotic}
\end{equation}

\begin{figure*}[ht!]
 \centering
\includegraphics[width=1\linewidth]{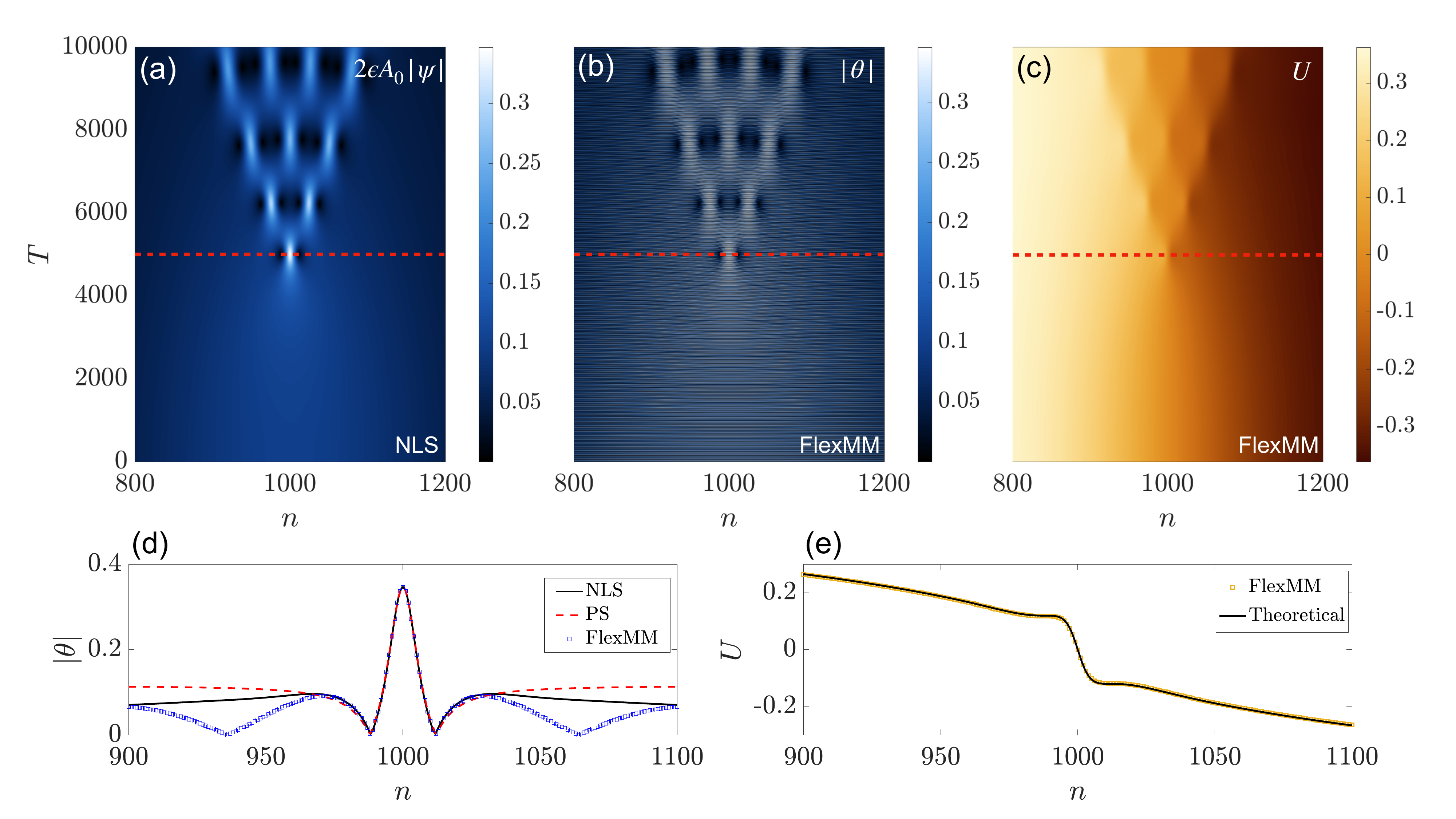}
 \caption{Numerical simulations of Eq.~(\ref{NLS_semi_classique}) are displayed in panel (a) and those of Eq.~(\ref{Full_discrete_chap_5}) are shown in panels (b-c) for an initial pulse with $N=10$. Panel (d) represents the absolute value of the rotational profile of the dynamics displayed in (a) and (b) at the maximum compression moment where the scaled PS is superimposed for comparison. In panel (e), the longitudinal displacement at the maximum compression moment (cf. panel (c)) is compared to the theoretical profile obtained using Eq.~(\ref{G0_chap_5}). This prediction is calculated by substituting the rotational field ($\psi$), computed with the NLS equation, into Eq.~(\ref{G0_chap_5}).}
\label{N_10}
\end{figure*}

Fig.~\ref{N_10} represents the dynamic evolution of the NLS equation (a) and of the discrete FlexMM equation (b-c) for $N=10$ in more details. The spatial profiles of $|\theta|$ (d) and $U$ (e) are represented at the maximum compression point. From the $|\theta|$ profile of the NLS model shown in panel (d), we can predict the profile of the first-order term of the longitudinal displacement $G_0$, using Eq.~(\ref{G0_chap_5}). This prediction is represented by the black line in panel (e). The excellent match between the predicted and actual FlexMM profiles demonstrates the effectiveness of the NLS model in accurately describing the envelope dynamics of the FlexMM for modulated waves of amplitude $A_0=5$.

\section{Effect of losses on the NLS and FLexMM maximum compression points}
\label{Sec:5.6}
The aim of the following section is to study the influence of system losses on the regularization of the gradient catastrophe phenomenon, particularly on the point of maximum compression. The linear losses parameters $\Gamma_{\theta/ U} $ are no longer assumed to be equal to zero. The initial conditions and parameters employed to solve Eqs.~(\ref{Full_discrete_chap_5}) and Eq.~(\ref{NLS_PQ_chap5}) are identical to those of the lossless configuration (see Sec.~\ref{sec:Numerical_NLS_chap_5}) except for the final time of integration which is now $T_f=2.10^4$. This approach is appropriate for the numerical resolution of the FlexMM equations of motion, as the losses parameters are assumed to be weak: $\Gamma_{\theta/U}$ are of the order $\mathcal{O}(\epsilon^2)$. 

\begin{figure}[!ht]
    \centering
\includegraphics[width=0.6\textwidth]{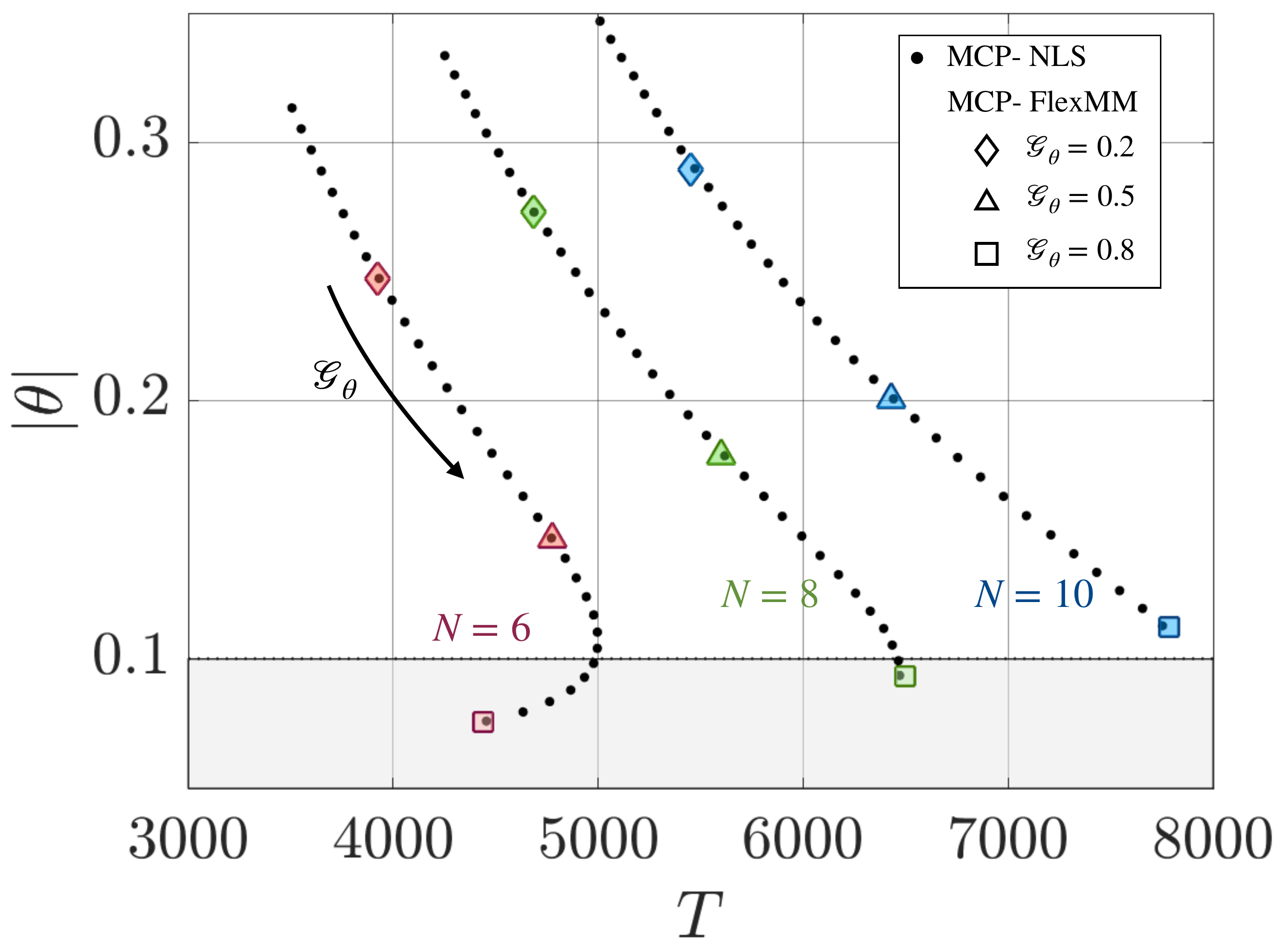}
    \caption{Evolution of the amplitude and moment of the maximum compression point (MCP) for various linear losses parameter $\mathcal{\mathcal{G}_\theta}$. It varied from 0 to 0.8 in steps of 0.025. The resulting MCPs predicted by the NLS equation are represented by black dots ($\bullet$). The arrow indicates the direction of increasing values of $\mathcal{G}_\theta$. MCPs obtained by the rotational motion of the FlexMM are depicted using diamonds ($\Diamond$) for $\mathcal{G}_\theta = 0.2$, triangles ($\triangle$)  for $\mathcal{G}_\theta = 0.5$, and squares ($\square$) for $\mathcal{G}_\theta = 0.8$. The study is conducted for $N=\{6,8,10\}$. The gray area indicates the amplitude range over which the maximum compression generated is lower than the initial amplitude.}
\label{MCP_eta}
\end{figure}

\begin{figure*}[ht!]
 \centering
\includegraphics[width=1\linewidth]{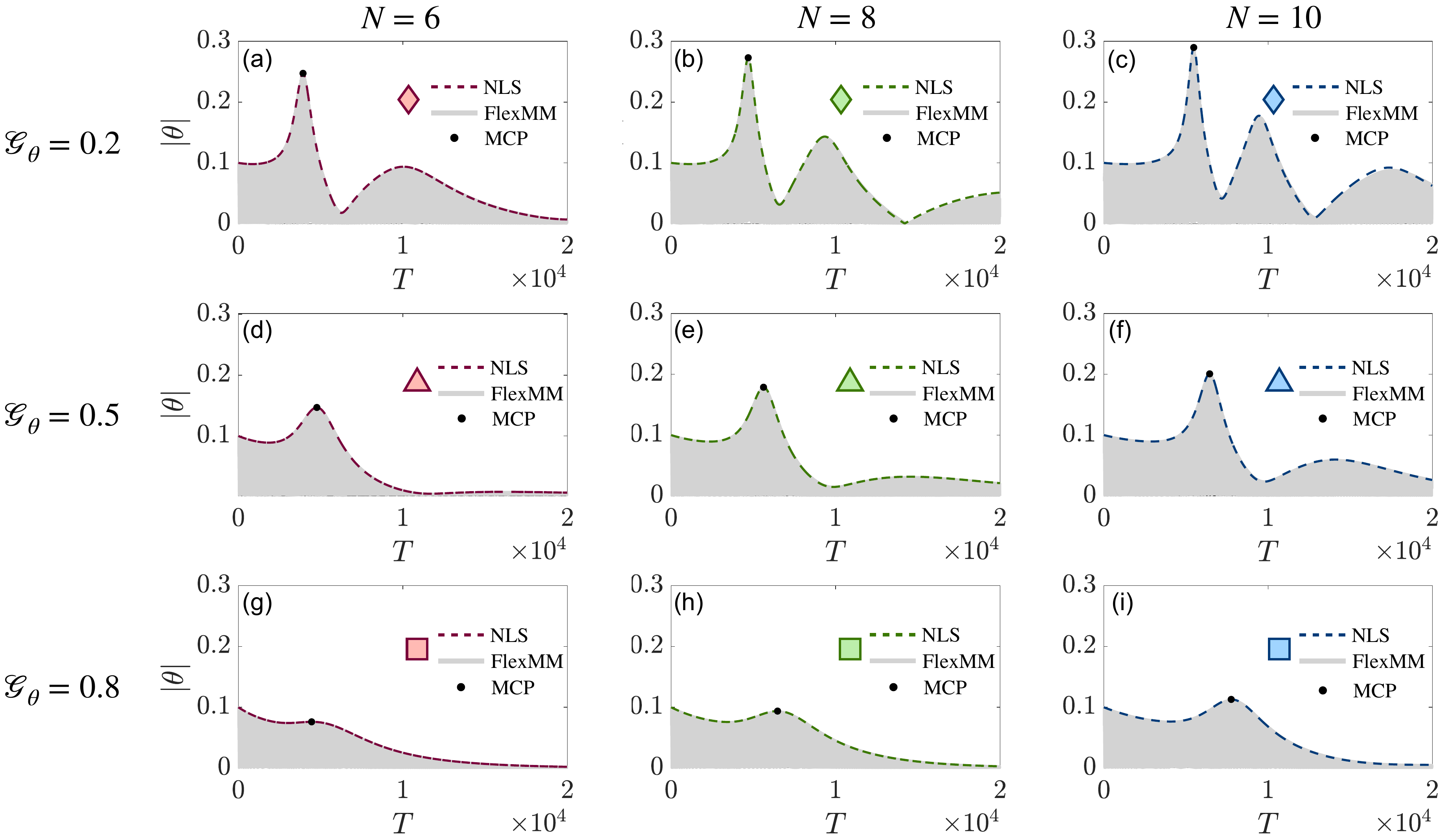}
\vspace{-0.7cm}
\caption{Temporal profile of the particle $n=n_s/2$ in the FlexMM (gray line) compared to the prediction from the effective NLS equation with linear losses (dashed line).  Each column represents simulations for a specific number of solitons, $N=\{6,8,10\}$, while each row corresponds to simulations for a specific linear losses parameter $\mathcal{G}_\theta=\{0.2,0.5,0.8\}$.} 
\label{Temporal_profiles}
\end{figure*}

In the results shown in Fig.~\ref{MCP_eta} and Fig.~\ref{Temporal_profiles}, the losses coefficients for both DOFs are chosen to be equal ($\Gamma_U = \Gamma_{\theta}$). Additionally, we observe that as long as  $\Gamma_U$ is of the same order as $\Gamma_\theta$ or smaller, the results exhibit no significant changes at the leading order for both DOFs, as predicted by the asymptotic expansion.

  Fig.~\ref{MCP_eta} compiles the characteristics of maximum compression points. The y-axis represents their amplitudes $|\theta|$ while the x-axis depicts their time of appearance $T$ for increasing values of $\mathcal{G}_\theta$ (see arrow) and for three soliton numbers $N=\{6,8,10\}$. Comparing to the FlexMM simulations, the results obtained using the NLS equation are represented in time $T$  (cf. Eq.~(\ref{change_of_variables_chap_5})) and for the scaled amplitude $|\theta| =2\epsilon A_0|\psi|$. We recall that the linear losses parameters of the NLS ($\mathcal{G}_\theta$) and of the lattice  ($\Gamma_\theta=\epsilon^2 \gamma_\theta$) equations are interconnected through Eq.~(\ref{mathcal_G_theta}). As observed for the lossless configuration in Fig.~\ref{Max_Amp_position}, the greater the $N$, the higher the amplitude of the structure generated by the gradient catastrophe regularization, and the later it appears. For a fixed value of $N$, the amplitude of the first localized structure decreases inversely to the losses coefficient increases, causing a delay in its appearance. For the $N=6$ branch and for the critical value $\mathcal{G}_{\theta c}=0.675$  we observe that the amplitude of the generated coherent structure is smaller than the initial amplitude. From this critical value, the first localized structure generated blends with the background amplitude and emerges earlier as $\mathcal{G}_\theta>\mathcal{G}_{\theta c}$ increases, see Fig.~\ref{Temporal_profiles}(g).

Fig.~\ref{Temporal_profiles} presents the temporal profile of the rotational DOF for the particle at the center of the chain ($n=n_s/2=1000$) along with the prediction provided by the NLS equation for some configurations used in Fig.~\ref{MCP_eta}. Overall, the results indicate that the dynamical evolution of the central site closely follows the NLS prediction, regardless of the values of $\mathcal{G}_\theta$ and $N$.

In the first row, panels (a–c), the system exhibits "weak" damping, particularly for $N=10$, due to the inverse proportionality $\gamma_\theta \propto N^{-1}$ (cf.~Eq.~(\ref{mathcal_G_theta})). As a result, see Fig.~\ref{Temporal_profiles}(c), other localized structures can appear around $T=10^4$ and $T=1.7.10^4$, before the total attenuation. As $\mathcal{G}_\theta$ increases, only one localized structure remains visible, as seen in panels (d-f). For larger values of $\mathcal{G}_\theta$ displayed in panels (g-i), these structures progressively merge and disappear, particularly when $N$ is small.

\section{Reliable estimation of extreme waves in a smaller lattice system}
\label{sec:smaller_lattice_system}
 From an experimental perspective, reducing the number of units composing the chain would be necessary. According to equation Eq.~(\ref{Le}), which defines $L_e$, the width of the initial envelope increases with $N$. Accordingly, to reduce the number of units required while maintaining a large $N$, we can play on both the initial condition (by increasing the amplitude $A_0$) and the parameters of the metamaterial itself ($\alpha$, $K_s$, $K_\theta$), to lower the $P/Q$ ratio.  However, it is important to note that as $A_0$ increases, the predictive accuracy of the NLS model for the evolution of the envelope $F_1$ decreases.
  \begin{figure*}[ht!]
 \centering
\includegraphics[width=1\linewidth]{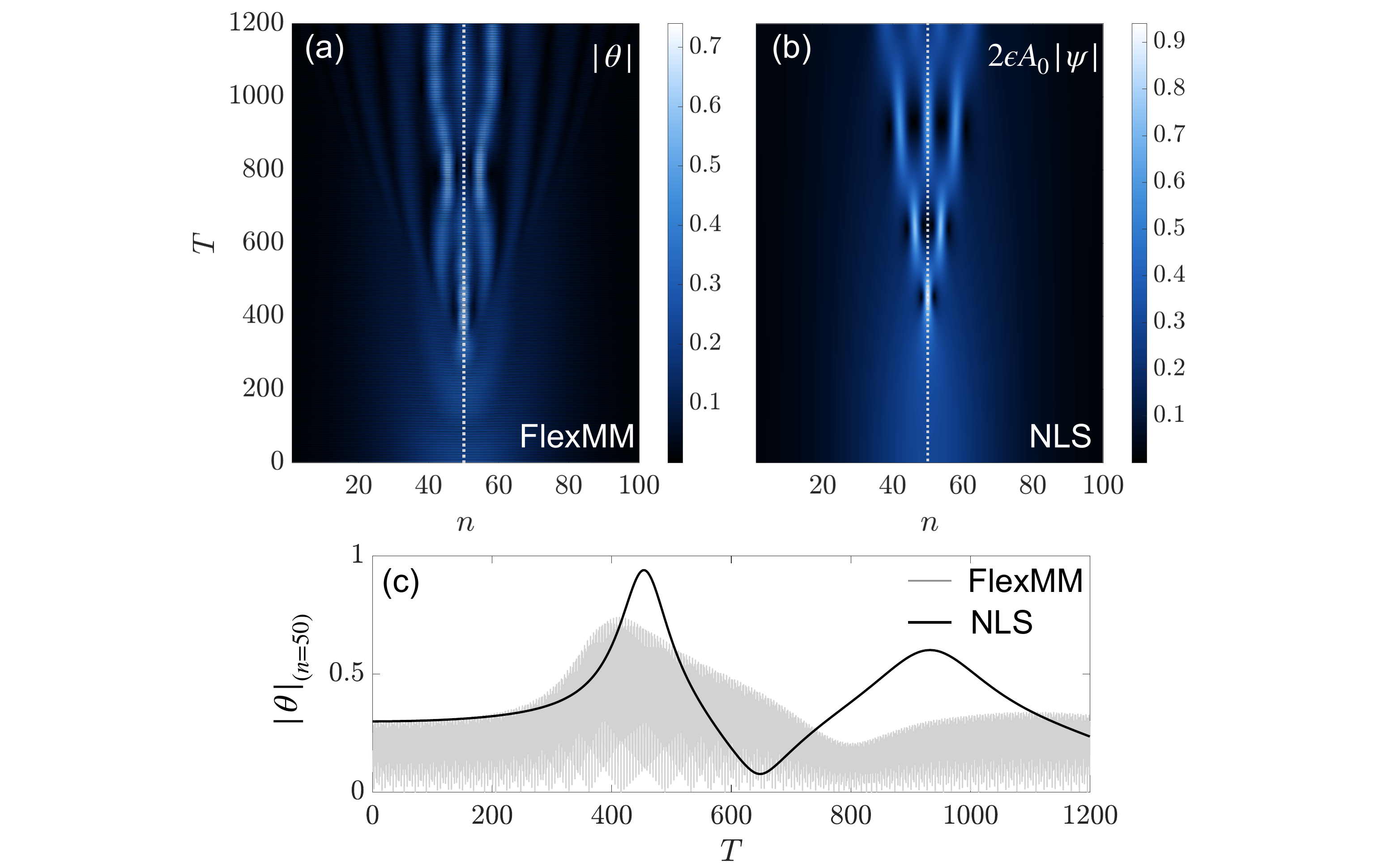}
\caption{Numerical simulations of Eq. (1) in panel (a) and Eq. (11) in panel (b) for an initial
pulse with $N = 6$. (c) Temporal profile of the unit $n = ns/2=50$ in the FlexMM (gray line), compared to the envelope evolution predicted by the effective NLS equation (black line).} 
 \label{smaller_lattice_no_losses}
\end{figure*}
 For the FlexMM parameters previously used in the simulations (detailed in Eq.~(\ref{parameters})), the ratio was $P/Q = 0.3270$. If $K_\theta=1.534\text{e}^{-1}$, the ratio decreases to $P/Q=0.0611$.

Fig.~\ref{smaller_lattice_no_losses} shows the effect of a lower $P/Q$ ratio combined with a higher initial amplitude set to $A_0 = 15$ in the lossless configuration $\Gamma_U=\Gamma_\theta = 0$. The spatio-temporal profile dynamics of the FlexMM (a) is compared to the scaled NLS predictions (b) while the temporal profile of $n=n_s/2$ displayed on panel (c) can be directly compared to the envelope predicted by the continuous NLS model. The results show that, based on NLS prediction, the time and amplitude of the first localized structure can be reasonably well estimated. However, once the extreme event has passed, the NLS model loses precision and even more, no longer accurately captures the rotational dynamics of the FlexMM.  Overall, with an appropriate choice of metamaterial parameters and initial pulse amplitude, extreme wave events can be well approximated even in lattice systems of experimentally feasible sizes, by a continuous nonlinear Schrödinger equation.
 
\section{Conclusions}
 In this work, we have shown how the gradient catastrophe phenomenon can manifest itself in FlexMMs through the propagation of weakly nonlinear and dissipative modulated waves, leading to the generation of the rational Peregrine soliton as a local asymptotic solution of the NLS equation. From the analytical model developed in \cite{demiquel_envelope_2024}, it has been proven that the rotational DOF evolution follows an effective NLS equation that incorporates a linear loss term while the longitudinal displacements are governed, at the leading order ($\epsilon$), by the dynamics induced by the nonlinear coupling through a dc-term. Based on the effective model predictions, we have shown that sufficiently large initial modulated waves with $k=0$ can give rise to new non-traveling 2-components envelope vector breathing soliton solutions of high amplitude compared to the background. Through a numerical study, both with and without system losses, we observed that this phenomenon persists across various losses parameters in the weakly dissipative regime.

In order to reduce the number of sites in the system for experimental feasibility and to preserve the accuracy of the NLS model, the physical parameters employed to build the FlexMM will need to be well chosen. Moreover, the formulation of our initial conditions problem needs to evolve to a boundary conditions one in order to realistically model the external mechanical driver that will be necessary for experiments. This driving force will induce a time-periodic forcing term in the effective NLS equation  \cite{diamantidis_exciting_2021} \cite{togueu_motcheyo_nonlinear_2024}.

Among the challenges related to structural design and fabrication, a promising research direction is to explore whether the onset of the gradient catastrophe can serve as a dynamic trigger for transition waves in bistable mechanical systems \cite{raney_stable_2016, paliovaios_transition_2024}, ultimately leading to structural reconfiguration. By initially setting all bistable unit cells in the same stable state, we hypothesize that the emergence of this extreme event at a controlled time and location will drive a targeted unit cell to its alternate equilibrium state, thereby possibly initiating a transition wave. This mechanism holds significant potential for applications in energy harvesting, mechanical switching, and adaptive mechanical architected materials.

\section*{Acknowledgement}
The authors acknowledge the support from the project ExFLEM ANR-21-CE30-0003-01.

\newpage
\newpage
\appendix
\section{Asymptotic expansion details}
\label{appendix_A}
In this appendix, we detail the procedure used to derive Eq.~(\ref{NLS_PQ_chap5}). Furthermore, this appendix demonstrates that the loss parameter $\gamma_u$ does not influence the solutions $G_0$ (order $\epsilon$) and $G_2$ (order $\epsilon^2$) of the longitudinal displacement $U$. Consequently, it does not affect the effective nonlinear Schrödinger (NLS) equation through the coefficients $P$ and $Q$.

First, substituting Eq.~(\ref{perturbative_expansion_chap_5}) into Eq.~(\ref{Motion_equation_Order_3_NL})(a) we arrive at the following equations collecting the dc in Eq.~(\ref{sub:exponential_dep_0sigma}) and $e^{2i\sigma n}$ terms in Eq.~(\ref{sub:exponential_dep_2sigma}) respectively,

\begin{subequations}
\begin{align}
    &\epsilon\ddot{G}_{0,n}= \epsilon\left(G_{0,n-1}-2G_{0,n}+G_{0,n+1}\right)-\frac{\epsilon^2}{2}\left(|F_{1,n-1}|^2-|F_{1,n+1}|^2\right)-\epsilon^3 \gamma_u\dot{G_0}\, ,\label{sub:exponential_dep_0sigma}\\
\nonumber\\
  &\epsilon^2\left( \ddot{G}_{2,n}-4i\omega\dot{G}_{2,n}-4\omega^2G_{2,n}\right) =  \epsilon^2 \left(G_{2,n-1}e^{-2ik}-2G_{2,n}+G_{2,n+1}e^{2ik}\right)-\epsilon^2\frac{F_{1,n- 1}^2e^{- 2ik}-F_{1,n+1}^2e^{ 2ik}}{4}\nonumber\\
  &-\epsilon^4 \gamma_u\left(\dot{G}_{2,n}-2i\omega G_{2,n}\right)\,.\label{sub:exponential_dep_2sigma}
  \end{align}
  \label{sub:U}
\end{subequations}

Similarly, substituting Eq.~(\ref{perturbative_expansion_chap_5}) into Eq.~(\ref{Motion_equation_Order_3_NL})(b) we get the following equation collecting the $e^{i\sigma n}$ terms,

\begin{equation}
\begin{split}
&\epsilon\left[\ddot{F}_{1,n}-2i\omega\dot{F}_{1,n}-\omega^2F_{1,n}\right]=\epsilon\alpha^2(K_\theta-K_s)\left[F_{1,n-1}e^{-ik}+2F_{1,n}+F_{1,n+1}e^{ik}\right]-\epsilon 6\alpha^2K_\theta F_{1,n}\\
&+\epsilon^3 3\alpha^2(K_s-1)|F_{1,n}|^2F_{1,n}-\epsilon^3\frac{\alpha^2}{2}\left[{2F_{1,n}\left(|F_{1,n-1}|^2+|F_{1,n+1}|^2\right)}+{F_{1,n}^*(F_{1,n-1}^2e^{-2ik}+F_{1,n+1}^2e^{2ik})}\right]\\
&+\epsilon^3\frac{K_s \alpha^2}{2}\left[|F_{1,n-1}|^2F_{1,n-1}e^{-ik}+2|F_{1,n}|^2F_{1,n}+|F_{1,n +1}|^2F_{1,n +1}e^{ ik}\right]+\epsilon^3 \frac{K_s \alpha^2}{2}\left[{F_{1,n}^2\left(F_{1,n-1}^*e^{ ik}+F_{1,n+1}^*e^{- ik}\right)}\right.\\
&\left.+{2|F_{1,n}|^2\left(F_{1,n-1}e^{- ik}+ F_{1,n+1}e^{ ik}\right)}\right]-\epsilon^3 2\alpha^2 F_{1,n}^*\left({G_{2,n+1}e^{2ik}-G_{2,n-1}e^{-2ik}}\right)-\epsilon^2 2\alpha^2 F_{1,n}\left({G_{0,n+1}-G_{0,n-1}}\right)\\
&-\epsilon^3\alpha^2\gamma_\theta(\dot{F}_{1,n}-2i\omega F_{1,n})\,.
  \end{split}
 \label{sub:exponential_dep_1sigma}
\end{equation}

We now proceed under the assumption that the discrete functions,
\begin{equation}
W_n(T)=\{F_{1,n}(T),G_{0,n}(T),G_{2,n}(T)\}
\end{equation}
exhibit slow variations across both space and time. As a result, we adopt the continuum limit approximation, replacing these discrete functions $W_n(T)$ with $W(X_1,X_2,...,T_1,T_2,...)$. Here, the variables $X_i=\epsilon^i X^i$ and  $T_i=\epsilon^i T^i$ are defined as slow variables, indexed by $i=1,2,...$. It is important to note that, within this approximation, the slowly varying functions are no longer depending on the fast variables $n$ and $T$.

In addition $W_{n\pm1}$, is computed up to order $\epsilon^2$ using Taylor expansion,
\begin{equation}
  W_{n\pm1}  =  W\pm \epsilon \frac{\partial W}{\partial X_1 }\pm \epsilon^2 \frac{\partial W}{\partial X_2}+\frac{\epsilon^2}{2}\frac{\partial^2 W}{\partial X_1^2}+\mathcal{O}(\epsilon^3) \, ,
  \label{Taylor_expansion}
\end{equation}
and the time derivation as,
\begin{equation}
   \dot{W_n}=\frac{\partial W}{\partial T} = \epsilon \frac{\partial W}{\partial T_1}+  \epsilon^2 \frac{\partial W}{\partial T_2}+\mathcal{O}(\epsilon^3) \,.
   \label{Temporal}
\end{equation}
By substituting Eqs.~(\ref{Taylor_expansion}-\ref{Temporal}) into the system of Eqs.~(\ref{sub:U}-\ref{sub:exponential_dep_1sigma}), we derive a hierarchy of equations corresponding to successive orders in $\epsilon$. At the lowest order, represented by Eq.~(\ref{sub:exponential_dep_0sigma}) (associated with $\epsilon^3$), a relationship emerges between the dc component $G_0$ and the envelope of the modulated plane wave $F_1$.
\begin{equation}
    \left(\frac{\partial^2}{\partial T_1^2} -\frac{\partial^2}{\partial X_1^2}\right)G_0 = \frac{\partial |F_1|^2}{\partial X_1}.
    \label{G0_appendix}
\end{equation}

In Eq.~(\ref{sub:exponential_dep_2sigma}) the lowest order is analogous to $\epsilon^2$ and relates $G_2$ to $F_1$,
\begin{equation}
  G_2 = \frac{i\sin(2k)}{8\left(\sin^2(k)-\omega^2\right)}F_1^2.
  \label{G2_appendix}
\end{equation}

We now move to  Eq.~(\ref{sub:exponential_dep_1sigma}) where at order $\epsilon^1$, we recover the dispersion relation,
\begin{equation}
    \omega^2 = 4\alpha^2(K_s-K_\theta)\cos^2\left(\frac{k}{2}\right)+6\alpha^2 K_\theta \, ,
    \label{order_1}
\end{equation}
which corresponds to the branch of the rotational waves of the non-lossy discrete model cf.~Eq.~(\ref{omega_2:subeq2_chap_5}).
At order $\epsilon^2$ we obtain the solvability condition,
\begin{equation}
      \frac{\partial F_1}{\partial T_1}+v_g\frac{\partial F_1}{\partial X_1}=0\, ,
\end{equation}
where
\begin{equation}
    v_g =-\frac{\alpha^2(K_s-K_\theta)\sin(k)}{\omega}\, ,
\end{equation}
is the group velocity corresponding to Eq.~(\ref{order_1}). Up to this order $F_1$ is linear and not coupled to $G_0$, $G_2$. At the order $\epsilon^3$ we have contributions from all fields and nonlinearity, leading to,

\begin{equation}
\begin{split}
        D_1^2 F_1-2i\omega D_2 F_1=&-2i\alpha^2(K_s-K_\theta)\sin(k)D_{2X}F_1-\alpha^2(K_s-K_\theta)\cos(k)D_{1X}^2F_1\\
     &+\left[8K_s\alpha^2\cos^2\left(\frac{k}{2}\right)-\alpha^2(5+\cos(2k))\right] |F_1|^2F_1\\
     &-4i\alpha^2\sin(2k)F_1^*G_2-4\alpha^2 F_1 D_{1X} G_0+\epsilon^3i\alpha^2\gamma_\theta\omega F_1\,.
    \end{split}
    \label{order_3_appendix}
\end{equation}

This expression, Eq.(\ref{order_3_appendix}), can be simplified by introducing the variables $\xi_i=X_i-v_gT_i$ and $\tau_i=T_i$, which correspond to a reference frame moving at the group velocity. Through this transformation and the substitution of $G_2$ ($G_0$) by Eq.~(\ref{G2_appendix}) (Eq.~(\ref{G0_appendix})), Eq.(\ref{order_3_appendix}) reduces to a nonlinear Schrödinger (NLS) equation that includes a linear loss term.
\begin{equation}
    i\frac{\partial F_1}{\partial \tau_2}+P\frac{\partial^2 F_1}{\partial \xi_1^2}+Q|F_1|^2F_1 =-i\frac{\gamma_\theta \alpha^2}{2}F_1 \, ,
    \label{NLS}
\end{equation}
with $P$ and $Q$, the dispersion and nonlinear coefficients respectively given by Eq.~(\ref{PQ_chap_5}).

\section{Bright envelope vector soliton analytical expression}
\label{appendix_B}

\begin{equation}
              \theta_1(X,T)=  2  A_0 \text{sech} \left[\frac{\epsilon}{L_e} (X-X_0- v_g  T)\right]\cos[kx-\Omega T] \, .
          \label{Analy_theta_appendix_B}
\end{equation}
$\Omega$ is the angular frequency of  the carrier wave defined as $\Omega= \omega^{(2)}-\epsilon^2\frac{Q A_0^2}{2}$.

\begin{equation}
        U_0(X,T) =\frac{A_0^2 L_e}{v_g^2-1} \tanh\left[\frac{\epsilon (X-X_0)-\epsilon v_g T}{L_e}\right]\, .
        \label{Analy_U_appendix_B}
\end{equation}

 \newpage
\bibliography{References}
\bibliographystyle{unsrt}
\end{document}